\documentclass[
preprint-reprint,
twocolumn,
 amsmath,amssymb,
 aps,
]{revtex4-1}
\usepackage{graphicx}% Include figure files
\usepackage{dcolumn}% Align table columns on decimal point
\usepackage{bm}% bold math
\begin{document}
%\preprint{APS/123-QED}

\title{Preferential attachment mechanism of complex network growth: "rich-gets-richer" or "fit-gets-richer"?\\}% Force line breaks with \\
%\thanks{A footnote to the article title}%
%Preferential attachment mechanism of complex network growth results from broad fitness distribution
\author{Michael Golosovsky}
\email{michael.golosovsky@mail.huji.ac.il}
% \altaffiliation[Also at ]{Physics Department, XYZ University.}%Lines break automatically or can be forced with \\
%\author{Sorin Solomon}%
\affiliation{The Racah Institute of Physics, The Hebrew University of Jerusalem, 91904 Jerusalem, Israel\\
}%
\date{\today}% It is always \today, today,
             %  but any date may be explicitly specified

\begin{abstract}
We analyze  the growth models for complex networks including  preferential attachment (A.-L. Barabasi and R. Albert, Science 286, 509 (1999)) and fitness model (Caldarelli et al., Phys. Rev. Lett. 89, 258702 (2002)) and demonstrate that, under very general conditions,  these two models  yield  the same dynamic equation of  network growth, $\frac{dK}{dt}=A(t)(K+K_{0})$, where $A(t)$ is the aging constant, $K$ is the node's degree, and $K_{0}$ is the initial attractivity. Basing on this result, we show that the fitness model provides an underlying microscopic basis for the preferential attachment mechanism. This approach yields long-sought  explanation for the initial attractivity,  an elusive parameter which was left unexplained  within the framework of the preferential attachment model. We show that  $K_{0}$ is mainly determined by the width of the fitness distribution. The measurements of $K_{0}$ in many complex networks usually yield the same $K_{0}\sim 1$. This  empirical universality  can be traced to frequently occurring lognormal fitness distribution with the width $\sigma\approx 1$.
%\begin{description}
%\item[PACS numbers]01.75.+m, 02.50.Ey, 89.75.Fb, 89.75.Hc
%\end{description}
\end{abstract}
% PACS, the Physics and Astronomy
                             % Classification Scheme.
%Use showkeys class option if keyword
                              %display desired
\maketitle
%\nocite{*}
\section{Introduction}
\label{Sec:intro}

Power-law distributions   were  brought to attention of scientific community about a century ago \cite{Mitzenmacher2004,Newman2005,Clauset2009} and they made sharp contrast with previously known Gaussians. The intriguing question arouse- what is the  generative mechanism   of these weird distributions? The explanation  came in 1970s when de Solla Price suggested his cumulative advantage model which he developed by studying network of citations to scientific papers.\cite{Price1976}  This model can be captured as follows.\cite{Newman2010} Consider a network consisting of nodes and edges. New nodes appear with constant rate $\frac{dN}{dt}$ and extend $\sim c$ edges to other nodes.  The probability of a new node $i$ to attach to a node $j$ is
\begin{equation}
\Pi_{j}=\frac{K_{j}+K_{0}}{\sum_{l}(K_{l}+K_{0})},
%\Pi_{i}=A(t)(K_{i}+K_{0})
\label{BA}
\end{equation}
where $K_{j}$ is the target node's in-degree, i.e., the number of its incoming edges, and the sum is over all nodes.   The initial attractivity $K_{0}$  ensures that newly born nodes start to acquire edges immediately. Equation \ref{BA} is motivated by  the Gibrat's proportional law which explained the power-law distribution of wealth. The accumulated edges in Eq. \ref{BA} play the role of node's "capital" and new edges are "dividends".

Equation \ref{BA} yields a power-law degree distribution, $p(K)\sim K^{-\gamma}$   with the exponent
\begin{equation}
\gamma=2+\frac{K_{0}}{c}.
\label{exponent}
\end{equation}
Price didn't look for detailed comparison of Eq. \ref{exponent} with the data, it was sufficient for him that Eq. \ref{BA} yields the power-law distribution with $\gamma\geq 2$ which is very similar to well-documented Pareto distributions of bibliometric indicators captured by Lotka's, Bradford's, and Zipf's laws. In the absence of any clue, Price postulated $K_{0}=1$.

The Price's cumulative advantage model (Eqs. \ref{BA},\ref{exponent}) didn't spread beyond the information science community since citation network was the only complex network then known. However,  as a result of proliferation of digitized information in 1990s, a number of information, biological, and social complex networks came to the forefront of scientific research, many of them being characterized by the power-law degree distributions  with $\gamma\sim3$.\cite{Boccaletti2006,Caldarelli2007,Newman2010,Barabasi2015} These distributions are somewhat different from the power-law distributions with $\gamma\sim2$ occurring in those social phenomena which are not considered as complex networks (word occurrence in  texts, city sizes, wealth distribution, etc.). To account for the power-law degree distributions  with $\gamma\sim3$  in information and biological networks Barabasi and Albert  suggested the preferential attachment model \cite{AlbertBarabasi2002} which is very similar to but not identical with the Price's cumulative advantage model.  The core assumption of the Barabasi-Albert model  is that a newly born node $i$ attaches to an older node $j$ with  probability
\begin{equation}
\Pi_{j}\sim K_{j},
\label{BA0}
\end{equation}
where $K_{j}$ is the target node's degree for undirected networks and the  sum of the in- and out-degrees,  $K_{j}=K_{j}^{in}+K_{j}^{out}$, for directed networks. Newman \cite{Newman2010} showed that for the latter case Eq. \ref{BA0} can be mapped onto the Price's model. Indeed,  since statistical distribution of  out-degrees in most complex networks is  narrow,  for directed networks,  Eq. \ref{BA0}  can be written as  $\Pi_{j}\sim (K_{j}^{in}+c)$  where $c=\overline{K_{j}^{out}}$.  On another hand, Eq. \ref{BA} can be written as $\Pi_{j}\sim (K_{j}^{in}+K_{0})$. The above  equations are equivalent and $c$ plays the role of initial attractivity $K_{0}$.  It should be noted, however, that while Price's model postulates   $K_{0}=1$,  the Barabasi-Albert  model assumes $K_{0}=c$. Both  models generate  networks with the power-law degree distributions, but the exponents are different: $\gamma\geq2$ for the former and $\gamma=3$  for the latter model. The Barabasi-Albert preferential attachment model  better agrees with  measured power-law distributions in complex networks and that is why it became the paradigm for complex network research, much in the same way as the Ising model established itself  as a  paradigm for  studies of magnetism.  In particular, the preferential attachment model became the basis for developing  specific network growth models. On another hand, it has been serving as a platform for organisation of measurements to characterize complex networks. %In what follows we do not make distinction between the Price's and Barabasi-Albert approaches and relate to Eq. \ref{BA} with unspecified  $K_{0}$  as the preferential attachment model.

Numerous measurements of the  growth dynamics of  complex networks validated Eq. \ref{BA}, namely they established  a linear or close-to-linear dependence between the growth rate of node's degree, $dK/dt$, and its current degree, $K$.  In addition, these measurements yielded a very small initial attractivity, $K_{0}\sim 1$. The last finding poses a problem for the theory. On the one hand, most complex networks are characterized by the power-law distribution with $\gamma\sim 3$ and $c>>1$ and, according to Eq. \ref{exponent}, this implies  $K_{0}\sim c>>1$. On another hand,  the measured $K_{0}\sim 1$ is much smaller and inconsistent with that inferred from Eq. \ref{exponent}. Solution of this inconsistency  requires (i) thorough analysis of the measurement protocols aimed at verification of Eq. \ref{BA}, (ii) reevaluation of the theoretical derivation of Eqs. \ref{BA},\ref{exponent},\ref{BA0}, and (iii) a deeper understanding of the initial attractivity $K_{0}$ which is not defined within the preferential attachment model.

%\ref{Sec:measurement},\ref{Sec:prediction},\ref{Sec:fitness},
The aim of our study is the reevaluation of the preferential attachment mechanism and its derivatives in order to understand this elusive parameter- initial attractivity. This is done in Sections \ref{Sec:evaluation} - \ref{Sec:explanatory} while in Section \ref{Sec:caldarelli} we analyze another kind of the network growth model- the  fitness model of Caldarelli et al. \cite{Caldarelli2002}. Section \ref{Sec:broad} reports our original results. Our main finding is that the fitness model  provides a microscopic basis of the preferential attachment mechanism and explains the initial attractivity which is determined by the shape of the fitness distribution. We also demonstrate that small initial attractivity $K_{0}$ is consistent with the power-law degree distribution with $\gamma\sim 3$. Section \ref{Sec:discussion} discusses the common basis of the preferential attachment, fitness model, and recursive search.
\section{Evaluation of the preferential attachment model}
\label{Sec:evaluation}

In what follows we do not make distinction between the Price's and Barabasi-Albert approaches and relate to Eq. \ref{BA} with unspecified  $K_{0}$  as the preferential attachment model. The success of this model in explaining the seemingly universal power-law degree distribution in complex networks prompted several theoretical generalizations.
\begin{itemize}

\item \emph{Initial attractivity}.  Refs. \cite{Dorogovtsev2000a,Ghoshal2013} analyzed Eq. \ref{BA0} in which the attachment rule was modified to $\Pi_{j}\sim (K_{j}+K_{0})$, where  $K_{j}$ is the total degree and $K_{0}$ is an arbitrary  number. It was found that the power-law degree distribution is retained but its exponent is modified in accordance with Eq. \ref{exponent}.

\item \emph{Accelerated network growth}. While the original Barabasi-Albert model assumed that new nodes and edges appear at the same rate, Leskovec et al. \cite{Leskovec2007} found that in many networks the number of edges grows faster than the number of nodes, in such a way that these  networks shrink with time. Assuming that the  average number of outgoing edges per node increases exponentially with time, $c=c_{0}t^{\Theta}$, Ref. \cite{Barabasi2015}  found that the power-law degree distribution in these shrinking networks is retained but its exponent increases by $\Delta\gamma=\frac{2\Theta}{1-\Theta}$.

\item \emph{Link editing.} In the original Barabasi-Albert model the existing links can't be deleted, the links between old nodes can't be added, and the old nodes can't be removed. Although this is true for citation networks, other networks such as Wikipedia edits, allow link or node editing. Ghoshal et al. \cite{Ghoshal2013} analyzed   networks for which  link editing and node removal are allowed and showed that these processes act disruptively on network topology, although there is a wide range of parameters for which the power-law degree distribution is conserved.

\item \emph{Aging.} Equation \ref{BA} assumes that the attachment probability $\Pi_{j}$ does not depend on node's age. However, the common sense tells us that  $\Pi_{j}$ should decrease with increasing age of the target node, in such a way that recent nodes become more popular. Refs. \cite{Dorogovtsev2000,Hajra2006,Wang2008,Wu2014,Ostroumova2016} considered a general case of the  preferential attachment with aging
\begin{equation}
\Pi_{j}= \frac{A(t_{j})(K_{j}+K_{0})}{\sum_{l}A(t_{l})[K_{l}+K_{0}]},
\label{aging}
\end{equation}
where $t_{j}$ is the age of node $j$ and $A(t)$ is the aging function which is usually assumed to follow exponential or power-law dependence, $A(t)\propto 1/t^{\nu}$. In the latter case Dorogovtsev and Mendes \cite{Dorogovtsev2000} showed that the power-law degree distribution is retained only for $\nu<1$ while for $\nu>1$ the aging effect overcomes the preferential attachment and the degree distribution does not follow the power-law dependence.

\item \emph{Memory}. Although Eq. \ref{aging} accounts for aging, it attributes  equal weight to all edges, as if the attachment were a Markov process. However, the recent edges are usually more important than the older ones, namely, attachment process can have memory. Refs. \cite{Wang2008,Miller2012,Rosvall2014,Gleeson2014,Ostroumova2016,Mokryn2016,Golosovsky2017} considered  preferential attachment with memory and replaced Eq. \ref{aging}  by
 \begin{equation}
\Pi_{j}(t_{j})\propto\int_{0}^{t_{j}}A(t_{j}-\tau) k_{j}(\tau)d\tau,
\label{memory}
\end{equation}
where $\Delta K_{j}(\tau)=k_{j}\Delta \tau$  is the number of edges garnered by the node $j$ in the time window ($\tau,\tau+\delta\tau$), $t_{j}$ is the node's age, $A(t_{j}-\tau)$ is the memory kernel, and the integral is over all past edges that appeared during target node's life.

\item \emph{Nonlinear preferential attachment.} While  Eq. \ref{BA} assumes a linear relation between attachment probability $\Pi_{j}$ and node's degree $K_{j}$, Krapivsky and Redner \cite{Krapivsky2001} considered a general case of nonlinear preferential attachment
\begin{equation}
\Pi_{j}=\frac{(K_{j}+K_{0})^{1+\delta}}{\sum_{l}(K_{l}+K_{0})^{1+\delta}}
%\Pi_{j}=A(t)(K_{j}+K_{0})
\label{nonlinear}
\end{equation}
where $\delta\neq 0$. It was shown that  the power-law degree distribution is closely associated with the linear case, $\delta=0$. For the superlinear attachment, $\delta>0$,  the network becomes the hub-and-spoke or  "winner-takes-all", while  for the sublinear attachment, $\delta<0$, the network becomes a gel-like where every node is connected to all other nodes and degree distribution is the stretched exponential rather than power-law.   Subsequently, Krapivsky and Krioukov  \cite{Krapivsky2008} showed that for weak superlinear attachment, $0<\delta<<1$, there is a vast asymptotic regime for which the network retains its power-law degree distribution.
\end{itemize}

In summary, theoretical studies indicate that the preferential attachment mechanism is  plausibly robust. Namely, it generates complex networks with the power-law degree distribution and the exponent $2<\gamma\leq 3$  under the following conditions: the attachment probability is  linear or weakly nonlinear, aging is weak, and initial attractivity is positive and  small. These conditions are quite reasonable and that is why the preferential attachment model has been accepted as a most plausible generative mechanism of  growing complex networks with the power-law degree distribution.

\section{Measurements and Model Validation}
\label{Sec:measurement}

Refs. \cite{Leskovec2008,Mislove2013,Kunegis2013,Perc2014,Pham2015}  discuss the microscopic mechanisms of the evolution of growing networks  and the ways of their validation. In what follows we present a more specific overview with the focus on preferential attachment.
%Pham, Sheridan, and Shimodaira \cite{Sheridan2012,Pham2015} developed a powerful numerical algorithm, based on maximum likelihood estimation,  to extract parameters of the preferential attachment from real data. In what follows we make an introduction

Straightforward verification of the preferential attachment model requires analysis of decisions made by incoming nodes. The measurements aimed at quantitative analysis of such decisions are widespread in psychology but they are rare in physics, biology, and computer science - the fields, where majority of complex networks appear. The most conventional way to uncover the growth mechanism of complex networks is to trace evolution of individual nodes.  To this end, the  perspective shall be shifted from the incoming node to  the target node. To perform such shift we go back to Eqs. \ref{BA}-\ref{nonlinear} and assume that new nodes appear at a constant rate. Consider $N$ new nodes that joined the network during time window ($t,t+\Delta t$). Each new node extends in average  $c$ edges to existing nodes. Consider one such target node $j$. From the  batch of new nodes,  it garners approximately $\Delta K_{j}=\Pi_{j}cN$  edges  where $\Pi_{j}$  is its attachment probability. We substitute there the general expression for $\Pi_{j}$, which includes nonlinearity, aging, and initial attractivity,  but not  memory,  and find
\begin{equation}
%\Delta K_{j}=\frac{A(t_{i})(K_{j}+K_{0})^{1+\delta}}{\sum_{j}A(t_{j})(K_{j}+K_{0})^{1+\delta}} c\Delta N
\Delta K_{j}=\tilde{A}(t_{j})(K_{j}+K_{0})^{1+\delta}
\label{BA1}
\end{equation}
where  $t_{j}$ is the target node's age at time $t$, $K_{j}$ is its current degree, and the aging function is  $\tilde{A}(t)=\frac{A(t)}{\sum_{l}A(t_{l})[K_{l}+K_{0}]^{1+\delta}}cN$. [Note difference between $A(t)$ and $\tilde{A}(t)$: for the Barabasi-Albert model $A(t)=1$ while $\tilde{A}(t)=\frac{2}{t}\frac{N}{\Delta t}$.] Equation \ref{BA1} is the basis for comparison of the preferential attachment model to measurements.

To validate Eq. \ref{BA1}  one usually considers a set of  all nodes of the same age and measures each one's degree  at  $t$ and at $t+\Delta t$. Then one calculates $\Delta K_{j}=K_{j}(t+\Delta t)-K_{j}(t)$, the number of additional edges that  each  node  garnered during the time window ($t,t+\Delta t$), plots $\Delta K_{j}$ versus $K_{j}$, and makes one's best to fit this scatter plot using Eq. \ref{BA1}.\cite{Redner2005,Mislove2013,Garavaglia2017}  This fit is by no means trivial. The catch here is that $\Delta K_{j}$ is a discrete  stochastic variable and Eq. \ref{BA1} predicts its  mean value but says nothing about the variance. Our measurements for citation networks \cite{Golosovsky2012} indicate that $\Delta K_{j}$ distribution (for fixed $K_{j}$) follows negative binomial distribution with high variance-to-mean ratio $>2$, in such a way that the variance of $\Delta K_{j}$  is considerably greater than that for the Poisson distribution. In other words, $\Delta K_{j}(K_{j})$ dependence is so noisy that direct fitting of $\Delta K_{j}$ vs $K_{j}$ using Eq. \ref{BA1} is not very informative.

To circumvent the problem of noise one can use logarithmic binning of $K_{j}$, plot a histogram of $\Delta K_{j}$, and find the trend. This method was originated by Newman \cite{Newman2001} and adopted by many others.\cite{Capocci2006,Massen2007,Kunegis2013,Golosovsky2013,Mislove2013,Higham2017,Higham2017a}

Another way to counter the noise problem is to plot  cumulative function $\int_{0}^{K} \Delta K(K)dK$ vs  $K$. In the context of complex networks  this procedure was first applied by Jeong et al. \cite{Jeong2003} and subsequently by Refs. \cite{Eom2011,Medo2014}. However, there is a pitfall here.  If $\Delta K_{j}$ were continuous variable with symmetric distribution, this cumulative procedure should certainly work. However, since $\Delta K_{j}$ is a non-negative discrete variable with highly skewed distribution, the cumulative procedure can distort  the results, in particular, when applied to validation of Eq. \ref{BA1}, it overestimates the initial attractivity.\cite{Golosovsky2013}

Yet another strategy is to use the raw $\Delta K_{j}$ versus $K_{j}$ plots and to apply sophisticated numerical fitting procedure to find parameters of Eq. \ref{BA1}.\cite{Kunegis2013} This procedure could probably become a valuable tool in complex network characterization.\cite{kunegis}

In what follows we present the summary of measurements aimed at validation of Eq. \ref{BA1}.
\begin{itemize}
\item \emph{Preferential attachment}. The growth of many complex networks does follow the preferential attachment model.\cite{Leskovec2008,Mislove2013,Kunegis2013,Perc2014,Pham2015} However, some of these networks exhibit preferential attachment only  for nodes with low and moderate degrees while the nodes with high degree exhibit anti-preferential attachment.\cite{Capocci2006,Kunegis2013}

\item \emph{Linear or nonlinear PA?} Early measurements claimed  linear or close to linear preferential attachment.\cite{Newman2001,Jeong2003} Later measurements using large datasets (citations to scientific papers \cite{Golosovsky2012,Higham2017a} and patent citations  \cite{Csardi2007,Higham2017}) revealed superlinear attachment with the exponent  $1+\delta\sim 1.25$. Social networks (scientific collaboration,\cite{Newman2001} movie actors \cite{Jeong2003,Eom2011}) exhibit sublinear preferential attachment with the exponent $1+\delta=0.8-0.9$.

 \item \emph{Aging function.}  Our measurements of citations to scientific papers \cite{Golosovsky2013} yielded $\tilde{A}(t)\sim(t-\Delta)^{-\nu}$, namely, a power-law decay with small delay $\Delta\sim$1-2 yr and the exponent $\nu=2$. Patent citations  yield a similar power-law aging function with $\nu=1.3-1.6$. \cite{Csardi2007,Higham2017} Zeng et al. \cite{Zeng2017} provide an overview of aging effects in citation networks.
     %Valverde et al. \cite{Valverde2007} in their studies of patent citations claim aging function following a Weibull-like form $\tilde{A}(t)\sim(t)^{\nu-1}e^{-(t/t_{0})^\nu}$ with $\nu=1.45$.

 \item \emph{Initial attractivity.} Early measurements of citations to scientific papers  were not statistically representative to make reliable estimate of $K_{0}$.\cite{Jeong2003,Redner2005} Subsequent studies of  patent citations   yielded small  $K_{0}\sim 1$.\cite{Csardi2007}  Our high statistics measurements of citations to scientific papers  \cite{Golosovsky2013} also yielded small $K_{0}\sim 1$. Eom and Fortunato \cite{Eom2011} analyzed network of citations between the APS (American Physical Society) journals  and found a bigger number,  $K_{0}\sim 7$ for younger papers and $K_{0}\sim 1-2$ for the papers that are at least 5 years old. (Note, however, that Ref. \cite{Eom2011} used cumulative procedure which is known to  overestimate $K_{0}$.\cite{Golosovsky2013}). Recent  studies of Higham et al. \cite{Higham2017,Higham2017a} yielded $K_{0}=1-1.8$ for patent citations and $K_{0}=1$ for Physical Review citations. Thus, all  measured initial attractivities are small, $K_{0}<<c$, and better conform to Price's conjecture, $K_{0}=1$, rather than to the Barabasi-Albert conjecture, $K_{0}=c$.
 \end{itemize}

Thus, preferential attachment mechanism of network growth, as captured by Eq. \ref{BA1}, has been qualitatively validated for  many complex networks. The attachment in most of them is  linear or close-to-linear, although deviations from the linearity are well documented. However,  Eq. \ref{exponent} is not supported by measurements. Indeed, the measurements indicate that initial attractivity $K_{0}$ is small.  In this case  Eq. \ref{exponent} yields the  power-law degree distribution with $\gamma\geq  2$ and this is in contrast to the power-law degree distribution with  $\gamma\sim 3$ observed in  majority of complex networks.\cite{Jeong2003,Csardi2007,Eom2011,Golosovsky2013}  This inconsistency  notwithstanding,  the preferential attachment model became a paradigm of complex network growth and a platform for network characterization.

\section{Specific predictions of the preferential attachment  model and their status versus measurements}
\label{Sec:prediction}
After scientific community became persuaded that the growth of  complex networks is accounted for by the preferential attachment model, the research  shifted  from the model validation to analysis of its predictions. Indeed, besides the power-law degree distribution, the complex networks generated using Eq. \ref{BA} should acquire  a very special structure.\cite{Newman2010,Barabasi2015}

\begin{itemize}
\item \emph{First mover advantage}. The preferential attachment model  predicts strong positive  correlation between the node's age and degree, namely,  the degree of the old nodes should be substantially higher than that of the recent nodes.  The measurements reveal such correlation but it is not strong and most new edges do not necessarily go to old nodes.\cite{Newman2009,Newman2014}
    %\cite{Adamic-Huberman}

\item \emph{Trajectory of the nodes of the same age}. The basic preferential attachment model predicts that the node's degree grows  with time according to the rule: $K_{j}(t_{j})=K_{0}\left[\left(1+\frac{t_{j}}{t}\right)^{\frac{c}{c+K_{0}}}-1\right]$  where $t$ is the age of the network at the moment when the node was born and $t_{j}$ is the node's age.\cite{Barabasi2015} Thus, the node's degree grows with time with deceleration and the trajectories of the nodes of the same age  should be very similar. However, the measurements show  that these trajectories   strongly diverge \cite{Kong2008,Golosovsky2012} and do not necessarily decelerate with time. In particular, citation networks demonstrate  many  "sleeping beauties" \cite{Ke2015} whose trajectories accelerate with time.

\item \emph{Degree distribution for the nodes of the same age}. According to the preferential attachment model, this distribution is narrow and close to exponential, $p(K)\sim K^{K_{0}-1}(1-t_{j}^{\frac{c}{c+K_{0}}})^{K}$.\cite{Newman2010,Barabasi2015} Measurements on Wikipedia  and citation networks showed that degree distribution for the nodes of the same age is much wider than exponential and is better described by the power-law or lognormal function.\cite{Huberman1999,Adamic2000,Redner2005,Kong2008,Golosovsky2017}

\item \emph {Degree-degree correlation}. Within the framework of the preferential attachment model the assortativity of the resulting network is determined by the initial attractivity $K_{0}$.\cite{Newman2001,Barabasi2015} In particular, for $K_{0}=c$ (the Barabasi-Albert model), the network shall be neutral, for $K_{0}>c$  it shall be assortative, and for $K_{0}<c$ (Price's model) it shall be disassortative.\cite{Barabasi2015} Direct measurements of the initial attractivity yield small  $K_{0}\sim 1$ implying that most networks should be disassortative.   While social networks are indeed disassortative, citation networks are not.\cite{Barabasi2015,Geng2009,Xie2016a,Sendina2016,Golosovsky2017} Thus, contrary to model prediction, there is no straightforward relation between the initial attractivity $K_{0}$ and the network assortativity.
    %Since the exponent $\gamma$ of the power-law degree distribution is also determined by $K_{0}$, there should be a strong correlation between the network assortativity  and $\gamma$.
\item \emph {Clustering coefficient}. The preferential attachment mechanism predicts that in large networks the clustering coefficient shall be vanishingly  small.\cite{Ostroumova2017} However, many real networks have high clustering coefficient.\cite{Leskovec2008}
\end{itemize}

Thus, several specific predictions of the preferential attachment model are inconsistent with measurements. This is unsurprising since these predictions were made assuming a basic version of the model, namely, linear preferential attachment and absence of aging (Eq. \ref{BA}). Since most studied networks exhibit weakly nonlinear attachment and strong aging, the proper  account of these two factors could modify some of the above predictions and make them  consistent with observations. However,  the problem of the wide degree distribution of the nodes of the same age and the paradox of the first-mover advantage can be hardly solved in such a way. To address these problems one needs to go beyond the framework of the preferential attachment model as captured by Eqs. \ref{BA}-\ref{BA1}.

\section{Fitness-based preferential attachment}
\label{Sec:fitness}

The difficulties associated with the application of the preferential attachment model and its derivatives for the quantitative account of  complex network growth call for alternative approaches. One such alternative is the attachment probability which  is proportional not to node's degree but to some other node's attribute such as  local  clustering coefficient,\cite{Bagrow2013}  node's rank,\cite{Fortunato2006} or PageRank coefficient.\cite{Zhou2016} However, the most popular alternative is the  Bianconi and Barabasi model \cite{Bianconi2001} that introduced fitness- an empirical parameter that characterizes the propensity of nodes to attract edges. The core assumption of the model is that the node's fitness is a constant number  and does not change with time.

\subsection{Multiplicative fitness- Bianconi-Barabasi model}
How fitness can be incorporated into dynamic equation of network growth? The Bianconi-Barabasi model \cite{Bianconi2001}  introduces  fitness  on top of the preferential attachment, namely, it postulates  that the attachment probability is the product of node's fitness $\eta$ and degree,
\begin{equation}
\Pi_{j}=\eta_{j}(K_{j}+K_{0}).
\label{BB}
\end{equation}
(To be consistent with  Eq. \ref{BA} we introduced here  initial attractivity $K_{0}$.)   Solution of Eq. \ref{BB} yields the node's trajectory $K_{j}(t)$. It turns out that this trajectory strongly depends on fitness: a  high fitness latecomer can outperform a low-fitness old node. Thus, fitness solves  the problem of the first-mover advantage. It also solves the problem of degree distribution for the nodes of the same age which is now determined by  fitness distribution  rather than by the acquired degree.

What is fitness? On the one hand, it  includes the notion of similarity  known as homophily in social networks.\cite{Bramoulle2012}  Indeed, complex networks are rarely uniform, they consist of communities and subcommunities. New nodes tend to attach to  similar nodes, namely, to those belonging to the same community.  To measure similarity one can use overlap of contents or  bibliographies for citation networks and WWW pages,\cite{Menczer2004,Ciotti2016}  or overlap of common neighbors in the general case.\cite{Newman2001,Liben-Nowell2003,Mislove2013} Another ingredient of fitness is  associated with quality or talent. This component   is not easy to estimate  when the  node first appears, it can be measured  only after it has garnered some edges. The obvious way to measure the Bianconi-Barabasi's fitness is through  Eq. \ref{BB}.  Thus, Kong et al.  \cite{Kong2008} studied the network of WWW internet pages, analyzed  trajectories of the pages of the same age, and successfully fitted them using Eq. \ref{BB}. The fitness turned out to be constant, as expected,  and the fitness distribution  turned out to be wide.

The most striking prediction of the Bianconi-Barabasi model is that for wide fitness distributions there are supercritical nodes that eventually take a lion share of edges.  Such supercritical nodes were  observed in citation networks \cite{Barabasi2012,Golosovsky2017a}  and this successful prediction became a reason of the wide popularity of the  Bianconi-Barabasi  model.\cite{Caldarelli2007,Ghadge2010,Wang2013,Carletti2015,Bell2017}
 %[The model does not solve the problem of assortativity but anyway it has been always  considered as a minor problem].

While Eq. \ref{BB} explains several  features of  complex networks, such as degree distribution of the nodes of the same age,  it  does not account for aging. To convert the Bianconi-Barabasi model into a quantitative tool that can be  be compared to measured node's trajectories, Wang, Song, and Barabasi  \cite{Wang2013}   replaced Eq. \ref{BB} by the following expression
\begin{equation}
\Pi_{j}=\eta_{j}A_{j}(t_{j})(K_{j}+K_{0}),
\label{WSB}
\end{equation}
where $A_{j}(t_{j})$ is the aging function, specific for each node, and $t_{j}$ is the node's age.  (Ref. \cite{Wang2013} denoted the aging function by $P_{j} (t)$  while we denote it by $A_{j}(t)$ to be consistent with Eq. \ref{aging}). The Wang-Song-Barabasi model (Eq. \ref{WSB})  builds upon the earlier approach of Ref. \cite{Medo2011} who introduced  node's relevance, $X_{j}(t)\sim \eta_{j}A_{j}(t)$.

Equation \ref{WSB} was validated using citation network of Physics papers covered by the APS database,\cite{Wang2013} whereas  the  aging function was approximated by the lognormal dependence $A_{j}(t_{j})=\frac{1}{\sqrt{2\pi}\sigma_{j} t_{j}}e^{{-\left(\frac{(\ln{t_{j}}-\mu_{j})^{2}}{2\sigma_{j}^{2}}\right )}}$
where $\mu_{j}$ and $\sigma_{j}$ are specific parameters for each node.   Pham, Sheridan, and Shimodaira \cite{Pham2015,Pham2016} developed a software package  based on Eq. \ref{WSB}  and demonstrated that it is a valid platform for quantitative description of the complex network growth.

This success notwithstanding, Eq. \ref{WSB} has several problems. First of all, it contains too many fitting parameters. In addition to dynamic node attributes (degree $K_{j}$ and  age $t_{j}$), the Wang-Song-Barabasi model adds another three   static attributes: $\eta_{j},\mu_{j}$, and $\sigma_{j}$. Secondly, the Wang-Song-Barabasi  and its parent  Bianconi-Barabasi model assume \emph{linear} preferential attachment.  This is an unlucky coincidence  that both these models were validated using citation networks which exhibit \emph{nonlinear} preferential attachment.\cite{Golosovsky2012,Golosovsky2013,Higham2017a}  While  the Wang-Song-Barabasi model can be  extended to account for nonlinearity, this extension requires an additional fitting parameter- attachment exponent- in such a way that the resulting model  becomes too sophisticated.

\subsection{Additive fitness}
The multiplicative fitness of the Bianconi-Barabasi model is not the only way it can be  introduced into dynamic equation of network growth.  Ref. \cite{Papadopoulos2012} introduced fitness through optimization procedure, while  Refs. \cite{Erguen2002,Menczer2004,Bedogne2006,Eom2011} introduced it additively, as follows:
\begin{equation}
\Pi_{j}\propto (K_{j}+\eta_{j}).
\label{RF}
\end{equation}
Equation \ref{RF} is nothing else but Eq. \ref{BA} where fitness $\eta_{j}$ replaces the initial attractivity $K_{0}$.

The growth dynamics described by Eqs. \ref{BB}, \ref{RF} are not that different as it could seem.  In fact, the combination of nonlinear preferential attachment (Eq. \ref{BA1}) with  additive fitness (Eq. \ref{RF})  mimics Eq. \ref{BB}, in particular, it   predicts supercritical nodes. To demonstrate this we adopt continuous approximation of Ref. \cite{Barabasi2015} and replace $\Delta K_{j}$ in Eq. \ref {BA1} by $\frac{dK_{j}}{dt}\Delta t$. In view of Eq. \ref{RF}  Eq. \ref{BA1} can be recast  as follows:
\begin{equation}
\frac{dK}{dt}=\tilde{A}(t)(K+\eta)^{1+\delta}
\label{BBB}
\end{equation}
where  index $j$ was dropped for brevity. We solve Eq. \ref{BBB} for $\delta>0$ and find
\begin{equation}
K(t)\frac{\eta}{\left[1-\delta\eta^{\delta}
\int_{0}^{t}\tilde{A}(\tau)d\tau\right]^{\frac{1}{\delta}}}-\eta.
\label{nonlinear-fitness}
\end{equation}
To analyze Eq. \ref{nonlinear-fitness} we assume for simplicity that the integral  $\int_{0}^{t}\tilde{A}(\tau)d\tau$ converges  as $t\rightarrow\infty$. We introduce $\eta_{crit}=\left[\delta\int_{0}^{\infty}
\tilde{A}(\tau)d\tau\right]^{-\frac{1}{\delta}}$, in such a way that  Eq. \ref{nonlinear-fitness} reduces to
\begin{equation}
K(t)=\frac{\eta}{\left[1-\left(\frac{\eta}{\eta_{crit}}\right)^{\delta}
\frac{\int_{0}^{t}\tilde{A}(\tau)d\tau}{\int_{0}^{\infty}\tilde{A}(\tau)d\tau}\right]^{\frac{1}{\delta}}}-\eta,
\label{nonlinear-fitness1}
\end{equation}
where $t$ is the node's age. For $\eta<\eta_{crit}$ Eq. \ref{nonlinear-fitness1} yields  $K(t)$ that increases with time and eventually achieves saturation, $K(\infty)=\frac{\eta}{\left[1-\left(\frac{\eta}{\eta_{crit}}
\right)^{\delta}\right]^{\frac{1}{\delta}}}-\eta$. However, for  $\eta\geq\eta_{crit}$, $K(t)$ does not achieve saturation, namely, the node's trajectory becomes supercritical. (In fact, it undergoes a finite-time singularity at certain $t_{0}$, in such a way that Eqs. \ref{nonlinear-fitness}, \ref{nonlinear-fitness1}  hold only for $t<t_{0}$.)  Thus, for the superlinear preferential attachment, $\delta>0$, Eq. \ref{RF}  predicts the supercritical nodes - exactly as Eq. \ref{BB} does.

\section{Explanatory models of network growth}
\label{Sec:explanatory}

The basic preferential attachment model and the fitness-based preferential attachment are not explanatory models, they lack realistic scenario explaining how the new node chooses the target nodes. Indeed,  Eqs. \ref{BA},\ref{BA1} imply that an incoming node shall know degrees of all  other nodes in the network in order to attach to some of them. This can be true for  collaboration and some other social networks,\cite{Centola2007,Centola2010} for which a new node is  familiar  with some limited set of nodes, but not for general networks for which global information on network connectivity is usually unavailable.\cite{prophecy}

The most popular explanatory model of  network growth is the recursive search \cite{Vazquez2001} also known as link copying or redirection,\cite{Krapivsky2005} random walk or local search,\cite{Jackson2007,Goldberg2015}  triple(triangle) formation,\cite{Wu2009}   triadic closure,\cite{Martin2013} or forest fire model.\cite{Leskovec2005,Subelj2013} The motivation for this class of models was the explanation of the high clustering coefficient commonly observed in complex networks. The recursive search mechanism assumes that a new node attaches to a randomly found node, explorers the network neighborhood of the latter, and with some probability attaches to one \cite{Jackson2007} or all \cite{Krapivsky2005,Lambiotte2016} of its ancestors. This scenario can include one-level \cite{Vazquez2001} or multilevel search,\cite{Leskovec2005,Lambiotte2007,Subelj2013} in the latter case a new node explores network vicinity of all previously chosen nodes. Vazquez \cite{Vazquez2001} showed that, in the absence of ageing and memory, the one-level recursive search mechanism results in the attachment probability
\begin{equation}
\Pi_{ij}= \lambda+ qK_{j},
\label{recursive}
\end{equation}
where $\lambda$ is the probability of random search, $qK_{j}$ is the probability of recursive search, and $K_{j}$ is the target node's degree. If we recast this equation as $\Pi_{ij}= q(K_{j}+\frac{\lambda}{q})$ we immediately realize that this is nothing else but  Eq. \ref{BA} with $K_{0}= \frac{\lambda}{q}$.

The above studies suggested only a generic scenario of the recursive search while to convert it into a quantitative model one needs to calibrate this scenario against the measurements. Recently, we performed such calibration using  citation networks \cite{Golosovsky2017}  and found that Eq. \ref{recursive} shall be supplemented with  aging, memory, and- most important- fitness. Namely, it turned out that the random search is not random but fitness-based.  Our findings imply the multilevel recursive search that develops according to the following scenario:  a new node $i$ performs a fitness-based search in the network, finds some target node $k$, and explorers its network neighborhood. It can choose a  nearest neighbor of $k$ as a new target node $j$. The  probability of such choice depends on the age of the parent node $k$ with respect to node $i$ (obsolescence or memory). Then the node $i$ explores a network neighborhood of a newly chosen node $j$ and so on. When we shift the perspective to the target node $j$,  this  scenario results in the following dynamic equation
\begin{equation}
\frac{dK_{j}}{dt}= \eta_{j}m(t)+ \int_{0}^{t}qe^{\gamma(\tau-t)}k_{j}(\tau)d\tau
\label{my-recursive}
\end{equation}
where $K_{j}$ is the target's node degree, $\eta_{j}$ is its fitness, $m(t)$ is the aging function, $q$ is the probability of recursive search,  $k_{j}(\tau)\delta \tau$ is the number of edges garnered by a node $j$ in the time window ($\tau$, $\tau+\delta\tau$), and $\gamma$ is the obsolescence coefficient.

In the limit  $\gamma{-1}<<t$  (short memory), Eq. \ref{my-recursive} reduces to %the main contribution to the integral in Eq. \ref{my-recursive} comes from  recent edges garnered between $t$ and  $t-1/\gamma$.  Thus, in the integral we can consider only small time interval $t-\tau<<t$.  We approximate $k_{j}(\tau)$ by $k_{j}(t)-(t-\tau) \frac{dk_{j}}{d\tau}|_{t}$, perform integration, and after some algebra arrive at
\begin{equation}
\frac{dK_{j}(t)}{dt}\approx \eta_{j} m(t)+\frac{q}{\gamma}\frac{dK_{j}(t-1/\gamma)}{dt}
\label{dynamics-short}
\end{equation}
Equation \ref{dynamics-short} is the first-order autoregressive model  where  time delay is $1/\gamma$ and $q/\gamma$ is the first-order autoregressive parameter.\cite{Golosovsky2012}   Similar models  were suggested by Refs. \cite{Simkin2007,Wang2008, Gleeson2014} under the name of preferential attachment with gradually-vanishing memory.

In the limit  $\gamma{-1}>>t$ (long memory),  Eq. \ref{my-recursive}  reduces to
\begin{equation}
\frac{dK_{j}}{dt}\approx \eta_{j} m(t)+qK_{j}(t)
\label{dynamics-Bass}
\end{equation}
This is nothing else but Eq. \ref{recursive} where the random-based search has been replaced by the fitness-based search. Equation \ref{dynamics-Bass} is also similar to Eq. \ref{BBB} which describes preferential attachment with additive fitness. A similar model was suggested by Menczer.\cite{Menczer2004}  Thus, fitness pops out as an important ingredient of the recursive search model as well.

\section{Fitness-only model}
\label{Sec:caldarelli}

Our analysis shows that all extensions and explanations of the preferential attachment mechanism eventually invoke fitness. While at the birth of the research field of complex networks  the node degree seemed to be the most important parameter  determining  growth dynamics,  subsequent studies focused more on node fitness.\cite{Pham2016} This prompts us to consider generative mechanisms where fitness rather than node's degree  plays the key role.

One such mechanism was suggested by Caladarelli et al.\cite{Caldarelli2002}  who  assumed that the  probability of attachment between  a new node $i$ and the target node $j$  is just $\Pi_{ij}=f(\eta_{i},\eta_{j})$  where  $\eta_{i}$ and $\eta_{j}$  are node fitnesses,  and $f(\eta_{i},\eta_{j})$ is the symmetric function of  its arguments (linking function). Ref. \cite{Caldarelli2002} considered additive linking function but the later publication of the same group \cite{Servedio2004} postulated multiplicative linking function, $f(\eta_{i},\eta_{j})\sim \eta_{i}\eta_{j}$. The latter assumption became more popular and it allows the following generalization. Consider a target node $j$.  If fitness is determined by similarity  and  all nodes belong to the same community, then the node $j$ will garner edges with the rate $\Delta K_{j}\propto \overline{\eta_{i}}\eta_{j}$ where $\overline{\eta_{i}}$ is the average fitness of incoming nodes. This average fitness can be absorbed into the aging function, in such a way that the probability of a new node $i$ to attach to existing node $j$ is
\begin{equation}
\Pi_{j}\sim \eta_{j}A(t_{j}).
\label{fitness}
\end{equation}
The fitness-only approach  captured by Eq. \ref{fitness}  was  developed further by Refs. \cite{Bedogne2006,Simkin2007,Ghadge2010,Nguyen2012,Luck2017}. Although it seems to represent a radical deviation from  the preferential attachment model, in what follows  we demonstrate that this is not so. Under very general conditions,  these two approaches yield the same dynamics of  network growth.  In particular,  we demonstrate that if the fitness distribution is  broad, then the fitness-based attachment (Eq.
\ref{fitness}) yields the same growth dynamics as the preferential attachment model (Eq. \ref{BA1}) with $\delta=0$. However, the  initial attractivity $K_{0}$ is not an arbitrary parameter anymore but is determined by the shape and width of the fitness distribution.

\section{Fitness model with broad fitness distribution mimics preferential attachment}
\label{Sec:broad}
Consider a directed acyclic network that grows according to  Eq. \ref{fitness}. We assign to each node a certain fitness $\eta$ drawn from some distribution $\rho(\eta)$ where  $\int_{0}^{\infty}\rho(\eta)d\eta=1$.  We further assume that the degree of each node grows according to an inhomogeneous Poisson process, in such a way that $\Delta K $, the  number of edges garnered by a node  during time window ($t,t+\Delta t$), is represented by the Poisson distribution $Poiss(\lambda,\Delta K)=\frac{\lambda^{\Delta K}}{\Delta K!}e^{-\lambda}$  with the rate
\begin{equation}
\lambda=\eta A(t)\Delta t,
\label{direct-PA}
\end{equation}
%For brevity, we dropped here and thereafter the index $j$.
where  $A(t)$ is the aging function, the same for all nodes, and the time $t$ is counted from the moment when the node joined the network.  Equation \ref{direct-PA} leaves some ambiguity in the definition of the fitness $\eta$ and aging constant $A(t)$ since it includes only their product.  To raise this ambiguity, we define the aging constant in such a way that $\int_{0}^{\infty}A(\tau)d\tau=1$. Under this constraint, the physical meaning of fitness $\eta$ is the long-time limit of node's degree, namely, $\eta\sim K(t\rightarrow\infty)$.

Since Eq. \ref{direct-PA} is memoryless, the number of edges that each node garners through the period from $t=0$ to $t$ also follows  Poisson distribution with the rate
\begin{equation}
\Lambda =\eta \int_{0}^{t}A(\tau)d\tau.
\label{direct-PA1}
\end{equation}

We focus on the ensemble of $N$ nodes that joined the network at the same time which we set as $t=0$. Among these nodes, the number of those that garnered $K$ edges by time $t$ is
\begin{equation}
N(K,t)=N\int_{0}^{\infty}\frac{\Lambda^{K}}{K!}e^{-\Lambda}\rho(\eta)d\eta,
\label{NK}
\end{equation}
where $\rho(\eta)$ is the fitness distribution and $\Lambda (\eta)$ dependence is given by Eq. \ref{direct-PA1}. During time window ($t, t+\Delta t$) each of these $N(K,t)$ nodes garners $\sim \lambda$ edges, in such a way that the  average number  of new edges garnered by each node from this subset is
\begin{equation}
\overline{\Delta K}= \frac{N\int_{0}^{\infty}\lambda\frac{\Lambda^K} {K!}e^{-\Lambda}\rho(\eta)d\eta}{N(K,t)}
\label{NK1}
\end{equation}
We substitute Eq. \ref{direct-PA} into Eq. \ref{NK1}, note that $\lambda=\Lambda \tilde{A}(t)$, where $\tilde{A}(t)=\frac{A(t)\Delta t}{\int_{0}^{t}A(\tau) d\tau}$, use the equality
\begin{equation}
\Lambda Poiss(\Lambda,K)=(K+1)Poiss(\Lambda,K+1),
\label{equality}
\end{equation}
and come to
\begin{equation}
\overline{\Delta K}|_{K}=\tilde{A}(t)(K+1)\frac{N(K+1,t)}{N(K,t)}.
\label{k}
\end{equation}
where $N(K+1,t)$ is the number of nodes that after time $t$ garnered  $K+1$ edges. If the fitness distribution is sufficiently broad, then $N(K+1)\approx N(K)$  for $K>>1$ (see Appendix) and  Eq. \ref{k} approaches asymptotically to
\begin{equation}
\overline{\Delta K}=\tilde{A}(t)(K+1).
\label{PA-wide}
\end{equation}
This expression is nothing else but the preferential attachment rule (Eq. \ref{BA1}) with  $\delta=1$ and $K_{0}=1$. A similar result was obtained earlier by Burrell \cite{Burrell2003} using a different approach. In the continuous approximation, Eq. \ref{PA-wide} reduces to $\frac{dK}{dt}=\tilde{A}(t)(K+1)$. This equation has uncanny resemblance to the famous expression for the photon emission rate for two-level atomic systems, $\frac{dN_{ph}}{dt}=Bn_{2}(N_{ph}+1)$, where $n_{2}$ is the number of atoms in the upper state, $N_{ph}$ is the number of photons, and  $B$ is the Einstein coefficient  for stimulated  emission. Thus, $K$ is the analog of $N_{ph}$ and $K_{0}$ is the analog of spontaneous emission.

\begin{figure}[ht]
\includegraphics*[width=0.4\textwidth]{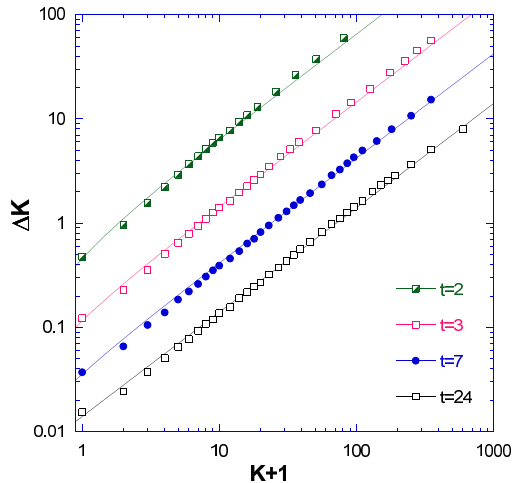}
\caption{ Numerical simulation of the growth dynamics of  400,000 nodes with the  aging function $A(t)=\frac{0.035t}{|t-2.4|^{1.3}}$ and lognormal fitness distribution  with $\mu=1.6$ and $\sigma=1.1$.  $\Delta K$ is the mean  growth rate, $K$ is the number of accumulated edges, $t$ is the age, and continuous lines show  fits to Eq. \ref{PA-wide1}. %(b) $\overline{\Delta K}$ versus $K$ for small $K$. The data for all $t$ lay on straight lines with common intercept $-1$.
}
\label{fig:PA}
\end{figure}

To validate  Eq. \ref{PA-wide}  through numerical simulation   we considered a set of 400,000 nodes  and simulated their growth using Eq. \ref{direct-PA}. We assumed a lognormal fitness distribution $\rho(\eta)=\frac{1}{\sqrt{2\pi}\sigma \eta}e^{{-\frac{(\ln{\eta}-\mu)^{2}}{2\sigma^{2}}}}$ with $\mu=1.6$ and $\sigma=1.1$, and the aging function $A(t)=\frac{0.035t}{|t-2.4|^{1.3}}$. The time was run from $t=0$ to $t=25$ with steps $\Delta t=1$, in such a way that $\sum_{0}^{t=25}A(t)=1$.  For each node $j$ in this set  we determined $K_{j}(t)$,  the total number of  edges accumulated  after  time $t$, and $\Delta K_{j}(t)$, the number of additional edges gained at step $t+1$. For every  $t$ we grouped all nodes into 40 logarithmically-spaced bins, each bin containing the nodes with close values of  $K$. For each bin, we determined $\Delta K$-distribution  and found its  mean, $\overline{\Delta K}$. Figure \ref{fig:PA} plots  $\overline{\Delta K}$ versus $K+1$, as suggested by Eq. \ref{PA-wide}. It is clearly seen that  this equation fits the data fairly well for $K>>1$. The fit at small $K$ is less satisfactory but it can be improved by using Eq. \ref{PA-wide1} with $K_{0}$ as a fitting parameter. We obtain
$K_{0}=0.7, 0.8, 0.85$ and 1 for $t=2,3,7$ and 24, correspondingly. Thus, anyway $K_{0}\sim 1$.

\begin{figure}[ht]
\includegraphics*[width=0.4\textwidth]{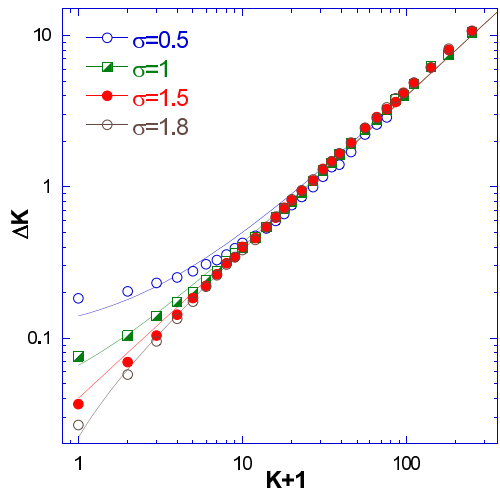}
\caption{Numerical simulation of the growth dynamics of the set of 400,000 nodes with different lognormal fitness distributions having the same $\mu$=1.6  and different $\sigma$. The symbols show results of numerical simulation, continuous lines show linear approximation $\overline{\Delta K}=A(K+K_{0})$ with $A=0.04$ and $K_{0}=3.5,1.65,1$, and 0.55 for $\sigma=0.5,1,1.5$,
and 1.8, correspondingly. The aging function is the same as in Fig. \ref{fig:PA}, $t=24$.
}
\label{fig:PA-sigma}
\end{figure}

\begin{figure}[ht]
\includegraphics*[width=0.4\textwidth]{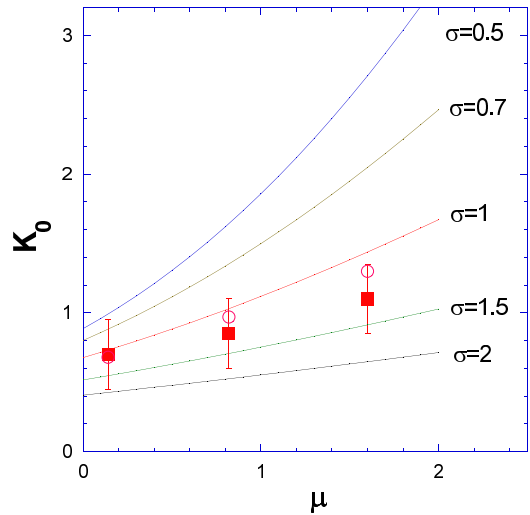}
\caption{Initial attractivity $K_{0}$ calculated from Eq. \ref{k-interp} in dependence of the parameters of the lognormal fitness distribution, $\mu$ and $\sigma$.  The filled  squares show our measurements for Physics, Economics, and Mathematics papers published in 1984 (see Ref. \cite{Golosovsky2012}). The open circles show our expectations based on measured $\mu$ and $\sigma$ of the lognormal fitness distribution for these very datasets. For all three research fields the measured fitness distributions have $\sigma=1.1$ and different $\mu$. The measured values of $K_{0}$  are close to those predicted by Eq. \ref{k-interp}.
}
\label{fig:mu-sigma}
\end{figure}

Figure \ref{fig:PA-sigma} shows $\Delta K(K+1)$ dependences for lognormal fitness distributions with different $\sigma$. These dependences  deviate
from Eq. \ref{PA-wide} for small $K$ and we approximate them  as
\begin{equation}
\overline{\Delta K}=\tilde{A}(t)(K+K_{0}),
\label{PA-wide1}
\end{equation}
where $K_{0}$ is the fitting parameter.   Equation \ref{PA-wide1} indicates that at small $K$, $\overline{\Delta K}\rightarrow\tilde{A}(t)K_{0}$.  On another hand, Eq. \ref{k} yields
\begin{equation}
\overline{\Delta K}|_{K=0}=\tilde{A}(t)\frac{N(1,t)}{N(0,t)}.
\label{k equal 0}
\end{equation}
Thus,
\begin{equation}
K_{0}\approx\frac{N(1,t)}{N(0,t)}
=\frac{\int_{0}^{\infty}\Lambda e^{-\Lambda}\rho(\Lambda)d\Lambda}{\int_{0}^{\infty} e^{-\Lambda}\rho(\Lambda)d\Lambda},
\label{k-interp}
\end{equation}
where $\rho(\Lambda)$ follows the lognormal distribution with shifted  $\mu'=\mu+\log (\int_{0}^{t}A(\tau) d\tau)$. Since $\int_{0}^{t}A(\tau) d\tau)\rightarrow 1$ in the long time limit, the difference between $\mu$ and $\mu'$ becomes increasingly small at long $t$. Figure \ref{fig:mu-sigma} shows $K_{0}$ calculated according to Eq. \ref{k-interp}  as a function of $\mu$ and  $\sigma$. We observe that $K_{0}$ increases with $\mu$ and decreases with $\sigma$. These dependences can be captured by the approximate empirical expression
\begin{equation} K_{0}\approx\frac{e^{\frac{\mu}{1+\sigma}}}{(1+\sigma^{2})^{0.6}}.
\label{K0}
\end{equation}
For reasonable values of $\mu=0-2$ and $\sigma=1-2$, $K_{0}$ lies between 0.5 and 1.5. It is determined by $\sigma$, and, to a lesser extent, by $\mu$. This means the following: if  $K_{0}$  is measured using Eq. \ref{PA-wide1}  using extrapolation from large $K$, then  one always gets $K_{0}=1$. On another hand, since most fitness distributions are broad, the $K_{0}$ measured using Eq. \ref{PA-wide1} for small $K$ (as it is usually done in most studies) is not not exactly 1 but close to 1, somewhere between 0.5 and 1.5. Figure \ref{fig:mu-sigma} shows that for narrow fitness distribution,  $K_{0}$ can be higher.

We plot on Fig. \ref{fig:mu-sigma} the  measured values of $K_{0}$ which were inferred from our studies of citation dynamics of scientific papers published in 1984. We considered  three research fields: Physics, Economics, and Mathematics and found that the fitness distribution for all three fields is lognormal with different $\mu$ but the same, $\sigma=1.1$. The measured and calculated initial attractivities $K_{0}$ are in good agreement and are all close to 1.

Thus, our numerical simulation supports the preferential attachment (cumulative advantage) model with  initial attractivity $K_{0}\sim 1$, as it was postulated by de Solla Price.\cite{Price1976} The natural question arises- why  $K_{0}\sim 1$ is so widespread?  Figure \ref{fig:mu-sigma} shows that the corresponding fitness distribution shall have any $\mu$ between 0 and 2 but the width shall be  $\sigma\approx 1$.  Nguyen and Tran \cite{Nguyen2012} showed by numerical simulation that if the complex network with a lognormal fitness distribution grows according to Eq. \ref{fitness}, then the power-law degree distribution with the exponent $\gamma\sim 3$ appears only for $\sigma\approx 1$. This observation implies that the initial attractivity is coupled to the exponent of the degree distribution, in such a way that the universality of initial attractivity $K_{0}\sim 1$ in growing networks is a by-product of the ubiquity of power-law degree distributions with $\gamma\sim 3$.

\section{Discussion}
\label{Sec:discussion}

Our analysis shows that if  network growth is considered from the perspective of a target node and  is studied using the mean-field approximation, namely, by averaging over many similar nodes, one cannot distinguish between the preferential attachment and the fitness-only models - both of them yield  Eq. \ref{PA-wide1}. Thus, in all that concerns the mean-field network dynamics, preferential attachment is equivalent to fitness-only  model, in other words, the rich-gets-richer mechanism reduces to the fit-gets-richer mechanism.\cite{Pham2016}  This is surprising since these two models are based on different premises. The preferential attachment model assumes that all nodes are born equal, the inequality in their degree coming by chance. After this  inequality has been established, it is  amplified by the autocatalytic process  represented by Eq. \ref{BA}. In contrast, the fitness model and the fitness-based recursive search model  assume that the nodes are born unequal, each newly born node is endowed with a certain fitness. The latent inequality in fitnesses becomes evident only after the nodes have been developing for some time. Surprisingly, the two opposing assumptions underlying network growth- all nodes are born equal or  different-  result in the same  growth equation,  Eq. \ref{PA-wide1}.

This does not mean that the two models are equivalent. While the preferential attachment model does not specify the initial attractivity, the fitness-only model with aging explains it perfectly well- it is determined by the shape of the fitness distribution. With respect to the power-law degree distribution in complex networks: the preferential attachment relates its to the strategy by which the new node attaches to old nodes, while  the fitness model implies that this distribution is inherited from the fitness distribution.  The fitness model successfully explains the first-mover advantage, degree distribution for the nodes of the same age, different trajectories of the nodes of the same age, etc. However, this model does not account for the nonlinear dynamic growth rule that is observed in some networks. Most important, this model does not account for the network structure: it addresses neither clustering coefficient nor assortativity.

Although it could seem that the fitness-only model is more advantageous than the preferential attachment,  Eqs. \ref{BA},\ref{BA1} can still be valid since the preferential attachment is a structural rather than explanatory model.  Indeed, the relation $\Pi_{ij}\sim K_{j}$ does not imply that a new node $i$ crawls through the whole network in order to gain information about degrees of all other nodes $j$. What occurs in reality is that the network grows following some local  rule  and this rule becomes imprinted in the network  topology.  When the network growth is analyzed, the changes in topology are visible while the underlying microscopic growth rule is not. This feeds the illusion that  the growth dynamics is  determined by network topology while in reality the reverse is true. The challenge is to uncover the  microscopic rules of network growth that explain the resulting network  topology. This can't be done basing on Eq. \ref{PA-wide1} since too many mechanisms yield the very same equation. Only truly microscopic measurements, such as  studying network growth from the incoming node's perspective, measuring autocorrelation and memory effects, can differentiate between the different  models. Our opinion is that the recursive search mechanism (which includes the fitness model as a particular case) is a best candidate to account for growth dynamics of many complex networks.
%Although we showed  that the fitness model provides justification of the preferential attachment mechanism and explains the initial attractivity, it can hardly serve as an explanatory platform of network growth since it is too phenomenological and devoid of  specific details characterizing  real networks.

Does the recursive search mechanism exclude the genuine preferential attachment- namely, the algorithm whereby a new node finds well-connected older nodes and attaches to them?  It has been generally believed that the recursive search mechanism is one of realizations of this algorithm, since if a new node makes a  random choice among the neighbors of already chosen nodes, it has high probability of picking up a highly-connected node. We  demonstrate here (Eq. \ref{dynamics-Bass}) that this strategy works in a straightforward way only if the recursive search does not have memory. In reality, recursive search has rather short memory,\cite{Golosovsky2017} and it is not clear whether highly-connected nodes can be found by this simple strategy: random choice among the neighbors of already chosen nodes.  We found \cite{Golosovsky2017}  that the recursive search follows a more clever strategy: the search in the network neighborhood of the previously chosen nodes is not random but has preference for those neighbors that are connected to several already chosen nodes.   The cartoon picture of such strategy is as follows. Simple recursive search: if Alice is linked to Bob, and Bob is linked to Frank, there is a chance that Alice will link to Frank. Clever recursive search:  if Alice is  linked to Bob and Charlie, and  both of them are  linked to Frank, then Alice will link to Frank almost for sure. Thus if a new node identifies a target node in the network vicinity of two or more previously chosen nodes, the probability of attachment to such node exceeds the sum of probabilities per each path, namely, multiple paths interfere constructively, reinforcing one another. The synergetic interaction between the paths to the next-nearest neighbors ensures that a new node finds highly-connected nodes. This strategy of exploring next-nearest neighbors  can still be considered as local strategy, but in fact, it is one step towards global search and this is the way how classical preferential attachment emerges in the recursive search.

\section*{Acknowledgments}
I am grateful to Lev Muchnik for valuable discussions.
\section{Appendix}
To explore the limits of  approximation $K>>1$ for which $N(K+1)\approx N(K)$, we consider the ratio $\frac{N(K+1,t)}{N(K,t)}=\frac{\int_{0}^{\infty}\frac{\Lambda^{K+1}}{(K+1)!} e^{-\Lambda}\rho(\Lambda)d\Lambda}{\int_{0}^{\infty} \frac{\Lambda^{K}}{K!} e^{-\Lambda}\rho(\Lambda)d\Lambda}$ and note  that for the uniform distribution, $\rho(\Lambda)=const$, the relation $N(K)=N(K+1)$ is satisfied exactly for every $K$.  In order to study to what extent this relation holds for non-uniform  distributions, we assume  a lognormal distribution, $\rho(\Lambda)=\frac{1}{\sqrt{2\pi}\sigma \Lambda}e^{{-\frac{(\ln{\Lambda}-\mu)^{2}}{2\sigma^{2}}}}$.  Then $\frac{N(K+1,t)}{N(K,t)}=\frac{\int_{0}^{\infty}\frac{\Lambda^{K+1}}{(K+1)!} e^{-\Lambda}e^{{-\frac{(\ln{\Lambda}-\mu)^{2}}{2\sigma^{2}}}}d\Lambda} {\int_{0}^{\infty}\frac{\Lambda^{K}}{(K)!} e^{-\Lambda} e^{{-\frac{(\ln{\Lambda}-\mu)^{2}}{2\sigma^{2}}}}d\Lambda}$. The expression  $\Lambda^K e^{-\Lambda}$, when considered as a function of $\Lambda$, is a bell-shaped function with a peak at $\Lambda_{max}=K$, the width of the peak being $\Delta\Lambda\sim K^{\frac{1}{2}}$. The relative width of this peak is $\frac{\Delta\Lambda}{\Lambda_{max}}\approx \frac{1}{K^{\frac{1}{2}}}$  and for $K>>1$ it is much narrow than any lognormal distribution with $\sigma>1$. In this case, the lognormal function is almost constant across the peak of the function $\Lambda^{K} e^{-\Lambda}$ and we can replace it by its value at the peak, $\Lambda=\Lambda_{max}$. Since $\int_{0}^{\infty}\frac{\Lambda^{K}}{(K)!} e^{-\Lambda}=1$ then, for $K>>1$, $\frac{N(K+1,t)}{N(K,t)}\approx e^{{-\frac{(\ln (K+1) -\mu)^{2}-(\ln K-\mu)^{2}}{2\sigma^{2}}}}= e^{-\ln(1+\frac{1}{K})\frac{\ln K-\mu}{\sigma^{2}}}\approx
e^{-\frac{\ln K-\mu}{K\sigma^{2}}}$. Since $\frac{\ln{K}}{K}<<1$ for $K>>1$, then $\frac{N(K+1,t)}{N(K,t)}\approx 1$. Thus, the latter relation holds for $K>>1$ and $\sigma>1$.

\onecolumngrid
\bibliography{references_master}

\begin{thebibliography}{92}
\bibitem [{pro()}]{prophecy}%
  \BibitemOpen
  \href@noop {} {}\bibinfo {note} {With the proliferation of easily accessible
  informational databases such as Google Scholar, Scopus, ISI Web of Science,
  etc., global information on many complex networks became easily accessible.
  In particular, for citation networks, the incentive to cite a certain paper
  may indeed come from the number of previous citations. Thus, the preferential
  attachment model is becoming a self-fulfilling prophecy.}\BibitemShut {Stop}%
\end{thebibliography}
\end{document}